\documentclass[aps,pra,showpacs,twocolumn]{revtex4-1}
\usepackage{amsmath,amssymb}
\usepackage{multirow}
\usepackage{graphicx}
\usepackage{color}
\usepackage{lastpage}
\usepackage{fancyhdr}
\usepackage{ifthen}

\begin{document}

\title{Phase transition in finite density and temperature lattice QCD}
\author{Rui Wang$^{1}$,
Ying Chen$^{2,3}$,Ming Gong$^{2}$, Chuan Liu$^{4,8}$, Yu-Bin Liu$^{1}$,\\
 Zhao-Feng Liu$^{2}$, Jian-Ping Ma$^{5}$,
 Xiang-Fei Meng$^{6}$, Jian-Bo Zhang$^{7}$\\
(CLQCD Collaboration)\\
$^{1}$School of Physics, Nankai University, Tianjin 300071, China\\
$^{2}$Institute of High Energy Physics, Chinese Academy of Sciences, Beijing 100049, China\\
$^{3}$Theoretical Center for Science Facilities, Chinese Academy of Sciences, Beijing 100049, China\\
$^{4}$School of Physics and Center for High Energy Physics, Peking University, Beijing 100871, China\\
$^{5}$Insititute of Theoretical Physics, Chinese Academy of Sciences, Beijing 100080, China\\
$^{6}$National Supercomputer Center in Tianjin, Tianjin 300457, China\\
$^{7}$Department of Physics,Zhejiang University, Hangzhou, Zhejiang 310027, China\\
$^{8}$Collaborative Innovation Center of Quantum Matter, Beijing 100871, China}

\begin{abstract}
We investigate the behavior of the chiral condensate in lattice QCD at finite temperature and finite chemical potential.  The study was done using two flavors of light quarks and with a series of $\beta$ and $ma$ at the lattice size $24\times12^{2}\times6$. The calculation was done in the Taylar expansion formalism. We are able to calculate the first and second order derivatives of $\langle\bar{\psi}\psi\rangle$ in both isoscalar and isovector channels.
With the first derivatives being small, we find that the second derivatives are sizable close to the phase transition and the magnitude of $\bar{\psi}\psi$ decreases under the influence of finite chemical potential in both channels.
\end{abstract}
\maketitle
\thispagestyle{fancy}
\fancyhead{}
\fancyhead[L]{\ifthenelse{\value{page}=1}{Submitted to 'Chinese Physics C'}}

\section{INTRODUCTION}
Since the advent of lattice QCD in the seventies by K.G.~Wilson~\cite{YMCA}, it has been proved that the theory is extremely successful in the analysis of low-energy dynamics among mesons and baryons~\cite{YMCB,YMCC}. Combined with large-scaled computations on supercomputers, people have been investigating various non-perturbative qualities, such as hadron spectrum, chiral transitions, behavior of the glueballs, hadronic matrix elements, spatial momentum dependence and vector current correlation\cite{WLYX,YMCD}.

However, lattice QCD also suffers from some shortcomings. For example, it violates some of the important symmetries that the continuum theory acquires and which can only be restored in the continuum limit. One of the most important symmetry is the chiral symmetry. A well known no-go theorem due to Nielson and Ninomiya~\cite{YMCE} says that chiral symmetry has to be realized differently on the lattice. The most elegant way is known and the fermion matrix of the lattice theory has to satisfy the so-called Ginsparg-Wilson relation~\cite{YMCF}. One example in this category is the so-called overlap fermion. However, practical simulations of overlap fermion encounters other technical problems and is rather costly. A compromise to this problem is to use staggered fermion which preserves part of the continuum chiral symmetry~\cite{HDME} and
runs effectively on supercomputers.

In this letter, using staggered quarks, we study the chiral condensate $\langle\bar{\psi}\psi\rangle$ and the relevant chiral symmetry breaking. As is well-known. $\langle\bar{\psi}\psi\rangle$ is an important order parameter in the measurement of phase transition in lattice QCD~\cite{SJLA,SJLB,SJLC}. Basically, the quantity $\langle\bar{\psi}\psi\rangle$ exhibits a fast decrease around the critical point but placidity at other place. This enables us to investigate the phase transition and chiral properties in both low temperature and QGP phase. Simulations of lattice QCD at finite density encounters another well-known long-lasting problem--the sign problem--the name borrowed from condensed matter physics
where it also appears in the simulation of models with fermions. In this letter we will follow the strategy of
Taylor expansion method~\cite{TaylorExpansion} in which all physical quantities are expanded around $\mu=0$.

This paper is organized as follows. In the next section, we briefly review the derivation of
the derivatives for the chiral condensate.  In the third part, our numerical results for the chiral condensate
are presented and we will summarize and conclude in the fourth part.

\section{Taylor expansion}

When the chemical potential is present, the simulation encounters the infamous sign problem.
However, if the chemical potential is not too large, everything can be expanded into
a Taylor expansion in $(\mu/T)$. For the purpose of this letter,
we need the expansion for $\langle\bar{\psi}\psi\rangle$,
\begin{eqnarray}
  \frac{\langle\bar{\psi}\psi\rangle(\mu)}{T^{3}} &=& \left.\frac{\langle\bar{\psi}\psi\rangle(\mu)}{T^{3}}\right|_{\mu=0}
  +\left(\frac{\mu}{T}\right)
  \left.\frac{1}{T^{2}}\frac{\partial\langle\bar{\psi}\psi\rangle}{\partial\mu}\right|_{\mu=0}\nonumber\\
  &&\!\!\!\!\!\!\!\!\!+\left(\frac{\mu}{T}\right)^{2}
  \left.\frac{1}{2T}\left(\frac{\partial^{2}\langle\bar{\psi}\psi\rangle}{\partial\mu^{2}}\right)\right|_{\mu=0}
  +O\left[\left(\frac{\mu}{T}\right)^{3}\right]
\end{eqnarray}
The first order and second order responses in the above expansion can be computed using numerical simulation at vanishing chemical potential. Using this equation, we can investigate the behavior of $\langle\bar{\psi}\psi\rangle$ at a small but non-vanishing chemical potential.

The staggered quark fermion matrix is given by,
\begin{eqnarray}
 D(U,\hat{\mu}) &=& ma\delta_{n,m}+\frac{1}{2}\sum_{\sigma=x,y,z}\eta_{\sigma}(n)\nonumber\\[-6pt]
  &&{}\times[U_{\hat{\sigma}}(n)\delta_{n+\hat{\sigma},m}-U_{\hat{\sigma}}^{\dag}(n-\hat{\sigma})\delta_{n-\hat{\sigma},m}]\nonumber\\
  &&{}+\frac{1}{2}\eta_{t}(n)[U_{\hat{t}}(n)e^{\hat{\mu}}\delta_{n+\hat{t},m}-U_{\hat{t}}^{\dag}(n-\hat{t})\nonumber\\
  &&{}\times e^{-\hat{\mu}}\delta_{n-\hat{t},m}]\;,
\end{eqnarray}
where $m$ is the bare quark mass and $\eta_{n,\mu}$ is a parameter only depend on the parity.
The parameter $\hat{\mu}=\mu/(N_tT)$ designates the chemical potential.
It is switched on for each flavor of quarks. In our simulation, we have both $u$ and $d$ quarks whose chemical potential will be denoted by $\mu_{u}$ and $\mu_{d}$ respectively.
 The quark propagator is related to the Dirac operator $D[U;\hat{\mu}]$ in the background gauge field configuration $U$ as:
\begin{equation}
 g(\hat{\mu})=D(U, \hat{\mu})^{-1}
\end{equation}
In lattice QCD, chiral condensate can be written as follows:
\begin{equation}
 \langle\bar{\psi}\psi\rangle\equiv Re\langle G\rangle=\left\langle\frac{1}{2} Re \{\text{Tr}[g(\hat{\mu}_{u})]+\text{Tr}[g(\hat{\mu}_{d})]\}
 \right\rangle\;,
\end{equation}
where $Tr$ implies summing over all indices and $\langle\cdot\rangle$ indicates
averaging over the gauge field ensembles.

For  the observable $\langle G\rangle$, one can take the first and the second order derivatives
with respect to the reduced chemical potential $\hat{\mu}\equiv \mu/T$ and obtain
\begin{eqnarray}
 \frac{\partial \langle G\rangle}{\partial\hat{\mu}} &=& \left\langle\dot{G}
 +G\frac{\dot{\Delta}}{\Delta}\right\rangle\nonumber\\
 \frac{\partial^{2}\langle G\rangle}{\partial\hat{\mu}^{2}} &=& \left\langle\ddot{G}+2\dot{G}\frac{\dot{\Delta}}{\Delta}\right\rangle
 +\left\langle G\frac{\ddot{\Delta}}{\Delta}\right\rangle
 -\langle G\rangle\left\langle\frac{\ddot{\Delta}}{\Delta}\right\rangle
\end{eqnarray}
where $\Delta$ is the fermion determinant given by
\begin{equation}
 \Delta=\det(D(\hat{\mu}_{u}))\det(D(\hat{\mu}_{d}))\;,
\end{equation}
and a dot indicates the derivative with respect to $\hat{\mu}$.

It is interesting to investigate the response of the chiral condensate  to both the
isoscalar  chemical potential (the traditional finite density case) and its isovector counterpart
(the finite isospin density case).
For the isoscalar we set
\begin{equation}
 \hat{\mu}_{S}=\hat{\mu}_{u}=\hat{\mu}_{d}\;,
\end{equation}
while for the isovector case we choose
\begin{equation}
 \hat{\mu}_{V}=\hat{\mu}_{u}=-\hat{\mu}_{d}\;.
\end{equation}
We will call these two cases as isoscalar and isovector channel in the following
by calculating the corresponding derivatives of $G$ with respect to the corresponding chemical potential. It is known that the first order derivative for both the isoscalar case and the isovector case are zero,
so the second order derivatives become crucial in our calculation.

\section{NUMERICAL SIMULATION AND RESULTS}

The conventional Wilson plaquette action is used which is characterized by the parameter $\beta$.
The gauge field configurations are generated using conventional $R$-algorithm of Hybrid Monte Carlo
with molecular dynamics step $\delta=0.01$ and trajectory length of $50$. The size of the lattices are all $24\times12^{2}\times6$ and for each parameter set, $600$ gauge field configurations are obtained.
By scanning the temperature dependence of the Polyakov loop
the ratio $T/T_c$ could be determined.
This information, together with other simulation parameters are summarized in Table I.

\begin{table}
 \centering
\begin{tabular}{|cccc|}
\hline
\hline
 $ma$&$\beta$&$N_{conf}$&$T/T_{c}$\\
\hline
0.020&5.292&600&0.90\\
\hline
0.020&5.327&600&0.95\\
\hline
0.020&5.347&600&0.98\\
\hline
0.020&5.373&600&1.02\\
\hline
0.020&5.392&600&1.05\\
\hline
0.020&5.422&600&1.10\\
\hline
0.015&5.317&600&0.95\\
\hline
0.015&5.337&600&0.98\\
\hline
0.015&5.363&600&1.02\\
\hline
0.015&5.382&600&1.05\\
\hline
0.012&5.327&600&0.95\\
\hline
0.012&5.347&600&0.98\\
\hline
0.012&5.373&600&1.02\\
\hline
0.012&5.392&600&1.05\\
\hline
\hline
\end{tabular}
\caption{Simulation parameters used in this study. All lattices are
of the size $24^{2}\times12\times6$ and $N_{conf}$ stands for the number of configurations.}
\end{table}

\begin{table}
 \centering
\begin{tabular}{|ccccc|}
\hline
\hline
 $ma$&$\beta$&$\langle\bar{\psi}\psi\rangle$&$\frac{\partial^{2}\langle\bar{\psi}\psi\rangle}{\partial\hat{\mu}_{S}^{2}}$
 &$\frac{\partial^{2}\langle\bar{\psi}\psi\rangle}{\partial\hat{\mu}_{V}^{2}}$\\
\hline
0.020&5.292&0.367(13)&-1.71(12)&-1.96(18)\\
\hline
0.020&5.327&0.351(13)&-2.26(31)&-2.75(30)\\
\hline
0.020&5.347&0.325(28)&-5.73(42)&-6.21(34)\\
\hline
0.020&5.373&0.138(24)&-6.73(45)&-7.12(53)\\
\hline
0.020&5.392&0.129(10)&-2.27(19)&-2.25(22)\\
\hline
0.020&5.422&0.119(8)&-1.43(5)&-0.98(4)\\
\hline
0.015&5.317&0.367(26)&-2.81(31)&-2.79(29)\\
\hline
0.015&5.337&0.335(33)&-7.07(59)&-6.55(56)\\
\hline
0.015&5.363&0.144(28)&-6.39(36)&-5.92(34)\\
\hline
0.015&5.382&0.123(8)&-2.13(20)&-2.62(24)\\
\hline
0.012&5.327&0.377(21)&-2.97(26)&-2.83(28)\\
\hline
0.012&5.347&0.347(34)&-6.08(41)&-6.17(45)\\
\hline
0.012&5.373&0.148(28)&-5.68(35)&-5.67(32)\\
\hline
0.012&5.392&0.126(7)&-2.55(23)&-1.84(17)\\
\hline
\hline
\end{tabular}
\caption{The values of $\langle\bar{\psi}\psi\rangle$ and its second order derivatives in the  isoscalar and isovector channel.}
\end{table}
We have calculated the value of $\langle\bar{\psi}\psi\rangle$ and its second order derivatives to both $\hat{\mu}_S$ and $\hat{\mu}_V$ for all our data sets and the results are listed in Table II.
Temperature dependence of the second order derivatives
of $\langle\bar{\psi}\psi\rangle$ with respect to $\hat{\mu}_S$
and $\hat{\mu}_V$ are illustrated in Figure 1 at $ma=0.020$.
\begin{figure}[thbp]
 \centering
 \includegraphics[width=2.4in]{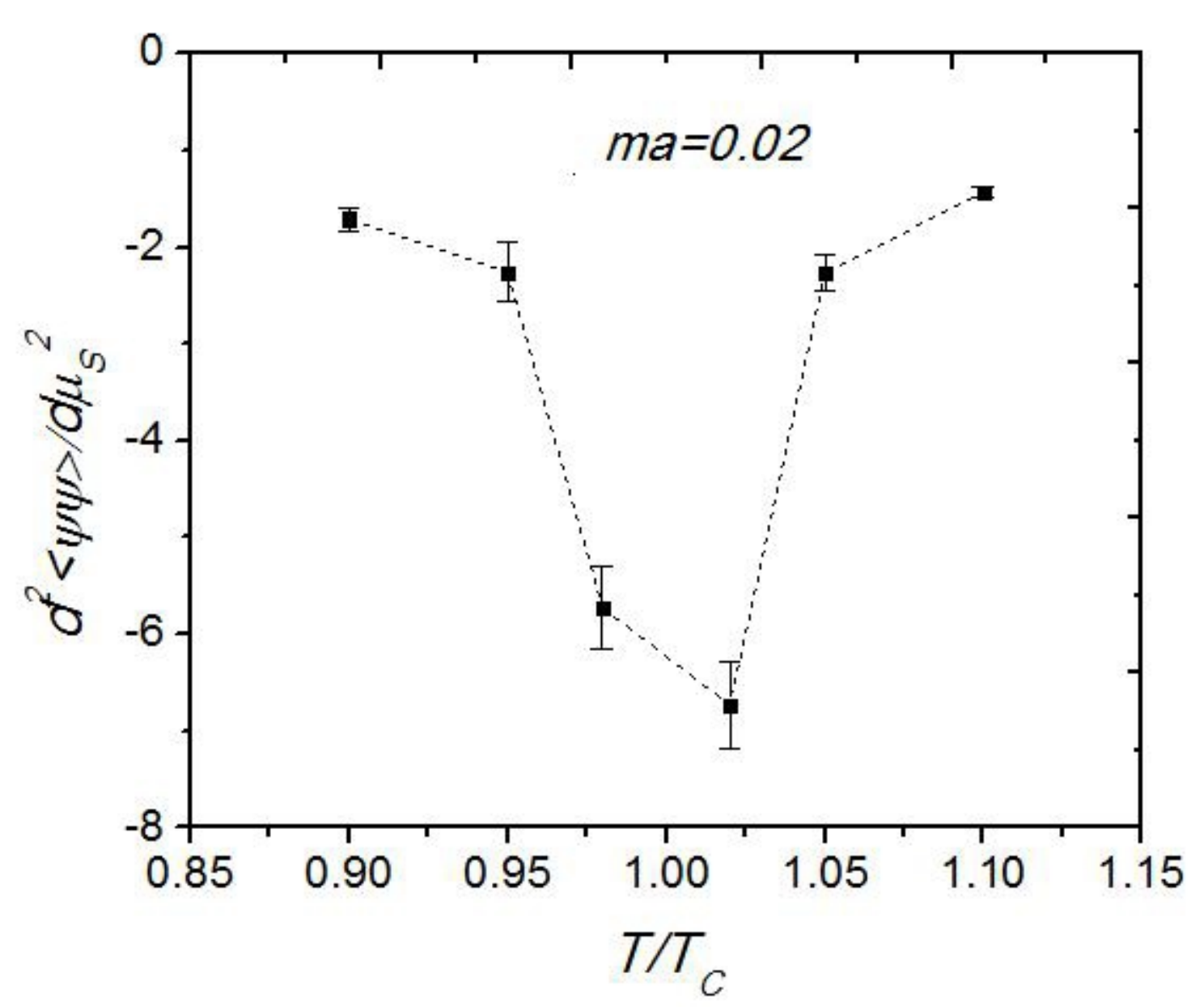}
 \includegraphics[width=2.4in]{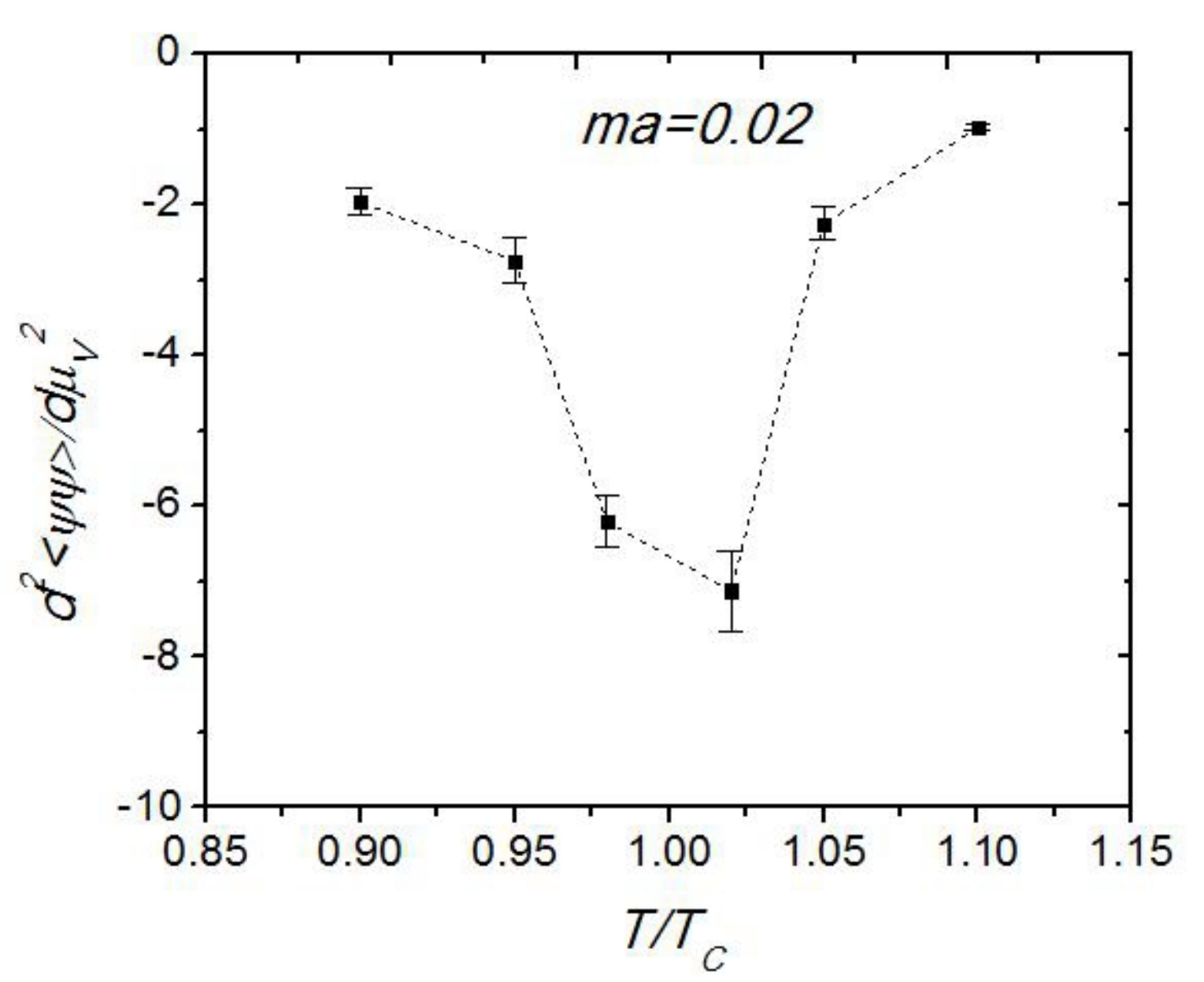}
 \caption{Second order derivatives of $\langle\bar{\psi}\psi\rangle$ at $ma=0.020$.}
 \label{fig.11}
\end{figure}
It is seen from the figure that the second order derivatives in both isoscalar and isovector channel share some common features. They are both negative but the absolute value is very small when away from the critical point. Close to the critical point, both increase substantially.

Now it is possible to consider the behavior of $\langle\bar{\psi}\psi\rangle$ with chemical potential and temperature in the critical region for both channels.  We omit the first order derivative for the quantity since they vanish and only consider the second order derivatives. Using the data listed in Table II, we obtain the following expansionw near $\beta_{c}$. For instance, slightly below $T_c$ at $\beta=5.347$ and $ma=0.020$, we obtain:
\begin{eqnarray}
 \frac{\langle\bar{\psi}\psi\rangle(\mu_{S})}{T^{3}}&=& 70.2(6.0)-17.2(1.3)\left(\frac{\mu_{S}}{T}\right)^{2}\nonumber\\
 &+&O\left[\left(\frac{\mu_{S}}{T}\right)^{3}\right]\\
 \frac{\langle\bar{\psi}\psi\rangle(\mu_{V})}{T^{3}}&=& 70.2(6.0)-18.6(1.0)\left(\frac{\mu_{V}}{T}\right)^{2}\nonumber\\
 &+&O\left[\left(\frac{\mu_{V}}{T}\right)^{3}\right]
\end{eqnarray}
As we see, the derivative corrections are quite substantial in the critical region.
Thus, the effect of the chemical potential makes $\beta_{c}$ to drop from its original value at $\mu=0$.
At an even lower temperature, for example, $\beta=5.292$ and $ma=0.020$, the data suggests
\begin{eqnarray}
 \frac{\langle\bar{\psi}\psi\rangle(\mu_{S})}{T^{3}} &=& 79.2(2.7)-5.13(37)\left(\frac{\mu_{S}}{T}\right)^{2}\nonumber\\
 &+&O\left[\left(\frac{\mu_{S}}{T}\right)^{3}\right]\\
 \frac{\langle\bar{\psi}\psi\rangle(\mu_{V})}{T^{3}} &=& 79.2(2.7)-5.88(55)\left(\frac{\mu_{V}}{T}\right)^{2}\nonumber\\
 &+&\left[\left(\frac{\mu_{V}}{T}\right)^{3}\right]
\end{eqnarray}
These derivative corrections are not that large as compared to the
case in the critical region.
In the phase above $T_c$ (in the QGP phase) at $\beta=5.422$ and $ma=0.020$, we obtain
\begin{eqnarray}
 \frac{\langle\bar{\psi}\psi\rangle(\mu_{S})}{T^{3}} &=& 25.8(1.8)-4.29(15)\left(\frac{\mu_{S}}{T}\right)^{2}\nonumber\\
 &+&O\left[\left(\frac{\mu_{S}}{T}\right)^{3}\right]\\
 \frac{\langle\bar{\psi}\psi\rangle(\mu_{V})}{T^{3}}&=& 25.8(1.8)-2.95(12)\left(\frac{\mu_{V}}{T}\right)^{2}\nonumber\\
 &+&O\left[\left(\frac{\mu_{V}}{T}\right)^{3}\right]
\end{eqnarray}

We can now plot the results for the chiral condensate at small chemical potential
in both the isoscalar and the isovector channel.
This is illustrated in Figure 2.
\begin{figure}[thbp]
 \centering
 \includegraphics[width=2.4in]{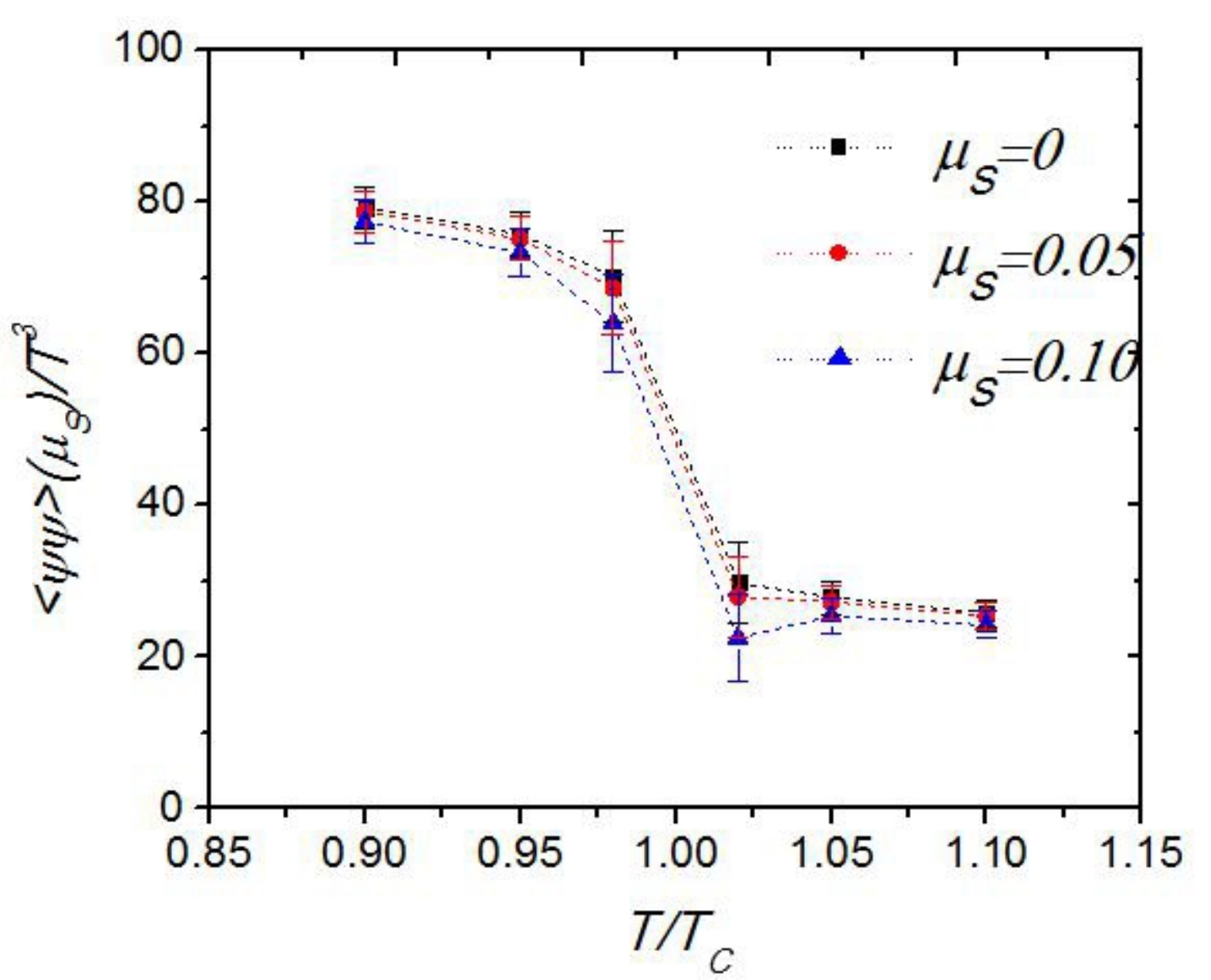}
 \includegraphics[width=2.4in]{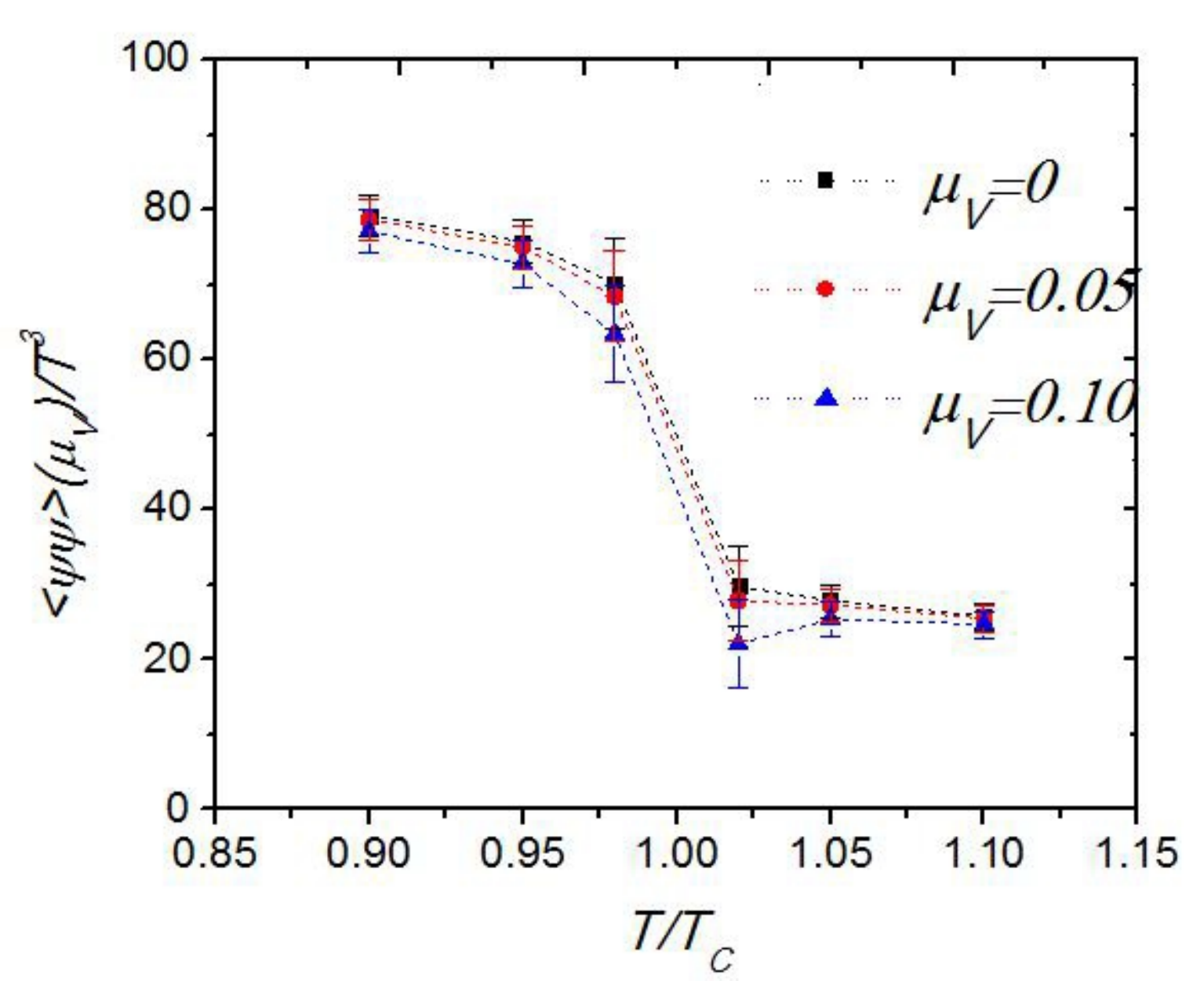}
 \caption{Behavior of $\langle\bar{\psi}\psi\rangle$ at finite isoscalar chemical potential (top panel) and the isovector potential (bottom panel).}
 \label{fig.11}
\end{figure}
In this figure, we include the behavior of $\langle\bar{\psi}\psi\rangle$ for the chemical potential $\hat{\mu}_{S,V}=0.05$ and $\hat{\mu}_{S,V}=0.10$. Since the second
order derivatives are all negative, we find the critical temperature tends to decrease under the influence of $\hat{\mu}_{S,V}$. Thus, in the low temperature phase, turning on the chemical potential brings the system closer to the phase transition where chiral symmetry is restored and decreases the magnitude of the chiral condensate.
In the high temperature phase, however, since chiral symmetry is already restored, the responses of
the chiral condensate to the isoscalar and isovector chemical potential are relatively small.

\section{CONCLUSIONS}

In this work, we have studied the response of the chiral condensate $\langle\bar{\psi}\psi\rangle$ to the chemical potentials using the
Taylor expansion method. The quantity is expanded around $\mu=0$ and the second derivatives of  $\langle\bar{\psi}\psi\rangle$ with respect to both $\mu_S$ and $\mu_V$ are obtained. As is seen,
though the first order derivatives are small, the second order responses are sizable and exhibit several unique features. The behavior of the responses for the $\langle\bar{\psi}\psi\rangle$ is closely related to the chiral restoration. For both isoscalar and isovector channel, we find that the critical temperature decreases with the influence of both chemical potentials, and the magnitude of $\langle\bar{\psi}\psi\rangle$ tends to decrease under the finite chemical potential.

\section*{ACKNOWLEDGMENTS}
This work is supported in part by the
 National Science Foundation of China (NSFC) under the project
 No.11335001. It is also supported in part by the DFG and the NSFC (No.11261130311) through funds provided to the Sino-Germen CRC 110 ``Symmetries and the Emergence of Structure in QCD''.
 The numerical calculations were performed on TianHe-1A supercomputer of the National Supercomputer Center in Tianjin.

\end{document}